\documentclass[a4paper,twoside]{article}
\newcommand{\affil}[1]{$^{\rm #1}$}
\usepackage[authoryear]{natbib}
\bibpunct{(}{)}{;}{a}{}{,}
\usepackage{graphicx}
\date{} 
\title{\large\bf\flushleft Examining alternatives to wavelet de-noising for astronomical source finding}
\author{\parbox{\textwidth}{\flushleft
\vspace{-0.5cm}
{\it Jurek, R.\affil{A,B}, Brown, S.\affil{A}}\\
\vspace{0.4cm}
{\small \affil{A}\,CSIRO Astronomy \& Space Sciences, Australia Telescope National Facility, P.O. Box 76, Epping, 1710 NSW, Australia.}\\
{\small \affil{B}\,Email: Russell.Jurek@CSIRO.au}}}
\begin{document}
\twocolumn[
\begin{changemargin}{.8cm}{.5cm}
\begin{minipage}{.9\textwidth}
\vspace{-1cm}
\maketitle
%
%
\small{\bf Abstract:}
The Square Kilometre Array and its pathfinders ASKAP and MeerKAT will produce prodigious amounts of data that necessitate automated source finding. The performance of automated source finders can be improved by pre-processing a dataset. In preparation for the WALLABY and DINGO surveys, we have used a test HI datacube constructed from actual Westerbork Telescope noise and WHISP HI galaxies to test the real world improvement of linear smoothing, the {\sc Duchamp} source finder's wavelet de-noising, iterative median smoothing and mathematical morphology subtraction, on intensity threshold source finding of spectral line datasets. To compare these pre-processing methods we have generated completeness-reliability performance curves for each method and a range of input parameters. We find that iterative median smoothing produces the best source finding results for ASKAP HI spectral line observations, but wavelet de-noising is a safer pre-processing technique. 

In this paper we also present our implementations of iterative median smoothing and mathematical morphology subtraction. 
 

\medskip
\medskip
\end{minipage}
\end{changemargin}
]
\small

\section{Introduction}
\label{intro}
Source finding reduces a dataset to a manageable abstract representation that is a collection of objects with physically meaningful properties. When a dataset becomes too large the dataset is virtually impossible to work with directly, and the catalogue is the only method of data exploration. This is the case for the two Australian Square Kilometre Array Pathfinder (ASKAP) HI surveys, Wide-field ASKAP Legacy L-band All-sky Blind surveY (WALLABY) (\citet{Koribalski_2009}; Koribalski, B., Staveley-Smith, L. et~al., in preparation) and Deep Investigation of Neutral Gas Origins (DINGO). Individual ASKAP spectral line observations will be at least 2,048 by 2,048 by 16,384 voxels, which is 256GB (512GB) in float (double) precision and only directly accessible using supercomputing facilities. 

The sheer size of the ASKAP spectral line observations combined with the number of observations required to carry out the WALLABY survey ($\sim$ 1200) necessitates automated source finding. An additional benefit is the reproducibility of automated source finders, which allows their performance to be incorporated into existing and future simulations of the WALLABY survey. The WALLABY team has been investigating existing source finding techniques as well as developing novel methods. The essential metrics for assessing automated source finding are reliability and completeness. Completeness is the fraction of sources that it recovers, and reliability is the fraction of detections that are actual sources. All automated source finders can be characterised by a `performance curve', which describes the combination of reliability and completeness that a source finder achieves on a given dataset. 

It is a common practise to pre-process a dataset before applying a source finding method. The goal of pre-processing is to improve both the completeness and reliability of the source finder. This is achieved by `correcting' the dataset. In a `corrected' dataset the noise behaves as your source finder assumes, the dataset is free from background structure and sources have maximised signal-to-noise ratios. 

It should be noted that the term `signal-to-noise ratio' in this context does not account for the Jy/beam units of radio observations. Technically a radio observation should be re-scaled for the new beam size when an observation is smoothed. This involves reversing the initial beam scaling, which in some circumstances increases the noise level more than it is minimised by smoothing. For the purposes of pre-processing and source finding though units are irrelevant. The signal-to-noise ratios discussed here therefore refer to the unscaled signal-to-noise ratios of radio observations, which are always enhanced by smoothing.

In this paper we will compare four pre-processing methods for ASKAP HI datacubes: iterative median smoothing, mathematical morphology subtraction, wavelet de-noising and linear smoothing. We will compare these pre-processing methods by examining the effect they have on the performance curve of a simple intensity threshold source finder. We analyse the effect on an intensity threshold performance curve, because intensity thresholding is at the core of most source finders eg. {\sc SExtractor} \citep{1996A&AS..117..393B} and {\sc SFind} \citep{2002AJ....123.1086H}. 

We are including linear smoothing and wavelet de-noising in our comparison, because they are among the most commonly used pre-processing methods. These two methods take contrasting approaches. Linear smoothing uses averaging or convolution to re-distribute the flux within the datacube so that noise fluctuations are reduced more than source signal, which results in increased source signal-to-noise ratios. Wavelet de-noising however tries to directly subtract noise from the datacube. A wavelet transform decomposes a datacube into signal on different scales at all positions within the datacube. Signal on scales smaller and larger than the expected size of sources can then be removed. Alternatively, the noise level on different scales can be measured from the wavelet transform, and only `significant' signal (as defined in some way by a user) on each scale is retained. We use the {\sc Duchamp} source finder \citep{2008glv..book..343W,Duchamp2} to implement both, because {\sc Duchamp} is not only a commonly used state-of-the-art source finder, but it is also the default ASKAP source finder. 

We are including iterative median smoothing in our comparison, because \citet{2009TAS.37.3.1172} has shown that iterative median smoothing produces a larger gain in source signal-to-noise ratio than linear smoothing methods. The key is that calculating a median is a non-linear process that preserves source `edges'. Source edges are preserved because median calculations are insensitive to sample outliers. Crucially, \citet{2009TAS.37.3.1172} found that only two iterations are required, so long as the first iteration uses the smallest smoothing kernel possible. This minimal number of iterations results in a reasonable computational load even when large smoothing kernels are used for the second iteration.

We chose to test mathematical morphology subtraction, because it is a proven technique for size filtering images. We can use mathematical morphology to filter out the small-scale information in a dataset to identify large scale structure in the image \citep{2002PASP..114..427R}. Subtracting this large scale structure can potentially improve reliability by re-normalising the dataset noise properties, so that the mean of the noise distribution is constant throughout the dataset.

There are distinct disadvantages common to all of these pre-processing methods. First, a poor choice of smoothing kernel can actually decrease source signal-to-noise ratios when using linear smoothing, iterative median smoothing and mathematical morphology subtraction. This is dealt with by using multiple smoothing kernels. This increases the computational load though, and the results of the multiple smoothing kernels need to be combined intelligently. Second, all of these pre-processing methods need to account for datasets having different types of dimensions. \citet{2011arXiv1112.3807F} is a good example of a wavelet transform for HI datacubes that accounts for the difference between the RA, Dec angular dimensions and the frequency dimension. 

There are two additional disadvantages of wavelet de-noising. Computing a wavelet transform can be much more computationally expensive than the other pre-processing methods. Additionally, a requirement of all wavelet transform kernels is that the integral of any kernel is zero. This prevents any individual wavelet transform kernel from matching an astronomical source, which is either a positive feature (emission) or a negative feature (absorption). Consequently, any astronomical source will necessarily exist on multiple scales (and probably multiple locations). Depending on how the wavelet transform information is filtered to de-noise a datacube, this can have a negative effect on source finder performance.

\begin{figure}[h]
\hspace*{-5mm}
\includegraphics[angle=270,width=0.55\textwidth]{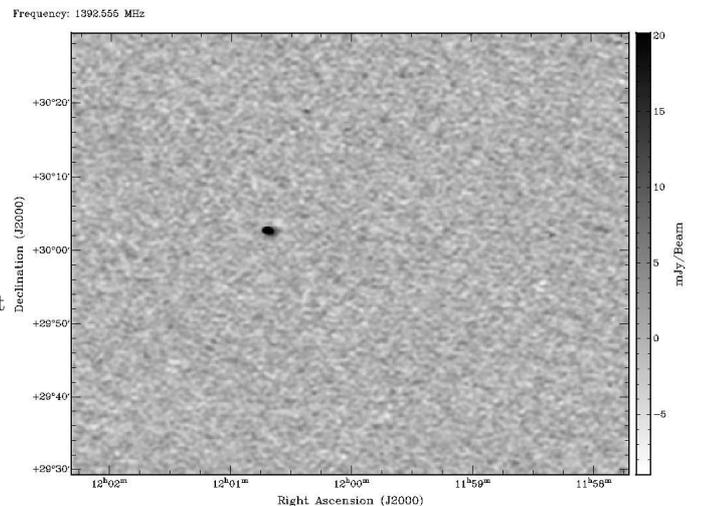}
\caption{This is a channel map of the WSRT test datacube used in this paper. The source in the centre of this channel map is one of the most spatially resolved sources.}
\label{cmap}
\end{figure}

We will carry out our comparison of these source finder pre-processing methods using the Westerbork Telescope (WSRT) test datacube in \citet{2011arXiv1112.3162S}. The WSRT test datacube was created by injecting WHISP sources \citep{2002ASPC..276...84V} into a datacube of real WSRT spectral datacube noise. This test datacube not only contains real sources embedded in real noise, but the resolution (both angular and spectral) and noise level closely matches that expected of the APERTIF and ASKAP telescopes. In particular, the 30$^{\prime\prime}$ Gaussian beam, 10$^{\prime\prime}$ pixels, 3.86 $km$ $s^{-1}$ channels and 1.86 $mJy/beam$ noise level of this test datacube, is designed to match WALLABY observations. This allows us to test the `real world' performance of the various pre-processing methods. We illustrate the scale of the WHISP sources and the test datacube using a channel map in Figure \ref{cmap}. 

The rest of this paper is organised in the following way. We begin by presenting the implementation of iterative median smoothing and mathematical morphology that we used for our comparison in Section \ref{ppimps}. Next we compare and analyse the performance impact of the various source finder pre-processing methods in Section \ref{analysis}. Then we finish in Section \ref{conclusion} with our conclusions and recommendations.

\section{Pre-processing implementations}
\label{ppimps}

\subsection{Iterative median smoothing}
Iterative median smoothing is reviewed and analysed in depth in \citet{2009TAS.37.3.1172}. Here we present a brief overview of iterative median smoothing and our implementation. Iterative median smoothing is the process of repeatedly replacing each element of a dataset with the median of a region centered on the element, using progressively larger regions. \citet{2009TAS.37.3.1172} found that with the right choice of region size, only two iterations are needed to obtain near maximal performance from iterative median smoothing. To do so, the first iteration needs to measure the median of the smallest region possible, and the second iteration needs to measure the median of a region matching the size and shape of the signal being optimised. The first pass removes elements that are outliers and the second pass re-distributes the flux, while preserving source edges, to improve source signal-to-noise ratio by averaging noise.

In this paper we use a two-pass implementation of iterative median smoothing. The first pass uses only the element being processed and the six neighbouring voxels that share a face with it. This is a 3-D extension of `four-connected' pixels. We chose a 3-D version of `four-connected' pixels for the first iteration, because this matches the pixel size of the beam's central component. This is sufficiently large to filter out individual noisy voxels in the presence of beam convolution (confirmed by us visually). The second pass uses either a rectangular parallelepiped or an ellipsoid extending along the frequency axis as a smoothing element. The ellipsoid (rectangular parallelepiped) is defined using separate radii (lengths) for the frequency axis and angular axes, RA and Dec.  

We have developed software that efficiently applies two-pass iterative median smoothing using an initial six-connected voxel element followed by a n-channel rectangular parallelepiped or ellipsoid element. This software deals with large datacubes using a two-pronged approach. First, the software uses a `buffer-and-shuffle' approach to minimise the memory overheads associated with multiply smoothing the input datacube. Second, sufficiently large datacubes are broken up into manageable `chunks', and processed sequentially. The use of a buffer-and-shuffle approach allows processing of files as large as three gigabytes on a 32-bit laptop before efficiency requires segmentation of a datacube.

\subsection{Mathematical morphology subtraction}
Mathematical morphology is a technique for analysing the morphology of objects in images. The core of mathematical morphology is erosion and dilation with a kernel. These two non-linear operations can be combined in multiple ways, but the simplest combinations are erosion followed by dilation to `open' an object and `closing' an object by dilating then eroding. The easiest way to think of the open and close operations is the effect that they have on text. The open operation sharpens the characters by filtering out small scale structure. A consequence of the open operation is that it `rounds' the remaining structure. The close operation by contrast blurs the characters. It amplifies small scale structure using the large scale structure as a guide. Unfortunately, sufficiently close characters will be merged into each other.

Monochromatic images, such as HI datacubes, are processed using `structuring element' kernels. Using a structuring element dilation is achieved by replacing an element with the maximum value in the region around it (specified by the kernel). Similarly, erosion is achieved by replacing an element with the minimum of the surrounding region. 

In this paper we use the approach developed in \citet{2002PASP..114..427R}. We use an open operation to filter the small scale structure out of the image and obtain an open image of the large scale structure. By subtracting this large scale structure from the original image, we can obtain a residual image of the small scale structure. We use the approach in \citet{2002PASP..114..427R} because combining the open and residual images preserves the flux of the original image.

\section{Analysis}
\label{analysis}
We compare the effectiveness of various source finder pre-processing methods by constructing completeness-reliability performance curves for a simple intensity threshold source finder, after applying the various pre-processing methods. We have chosen to compare the performance curves instead of completeness and reliability for a given threshold, because each pre-processing method will alter the datacube and its noise distribution in different ways. An arbitrary threshold (in units of noise level) will therefore not be consistent across the outputs of the various pre-processing methods.

The simple intensity threshold source finder that we use here is a two step process. In the first step a C++ code robustly measures the datacube's standard deviation from the interquartile range, and then selects all voxels greater than a user specified multiple of the standard deviation as source voxels. These flagged voxels are then combined into objects, merged, size filtered and turned into a catalogue using the object generating library presented in \citet{2011arXiv1112.1561J}. Every catalogue used merging lengths of 1, 1 and 3 empty voxels along the RA, Dec and frequency axes. We size filtered every catalogue to only include object's containing 14 voxels, occupying 5 lines of sight and whose extent in the RA, Dec and frequency dimensions is at least 3 voxels. We chose these merging and size filtering parameters, because they are representative of the values that would be chosen by a user when trying to maximise completeness.

The object generating library presented in \citet{2011arXiv1112.1561J} can generate catalogues that include a sparse representation of each object's voxel mask. Using the same custom C++ software as \citet{2012arXiv1201.3994P}, we use these sparse representations to match the object's in the various output catalogues to the input catalogue on a voxel-to-voxel basis. From these matches we calculate the source finder performance metrics.

\begin{figure}[h]
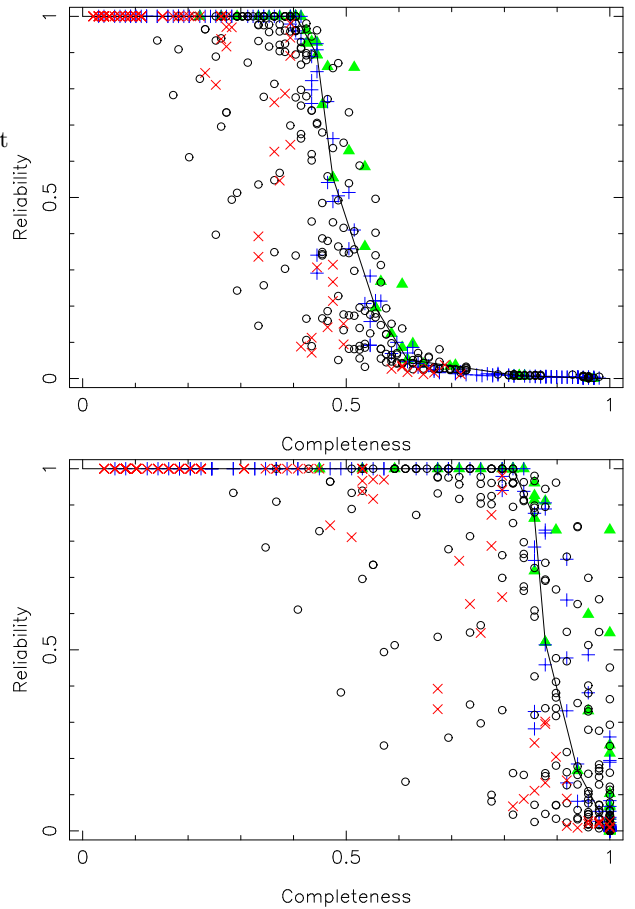

\hspace*{-6mm}
\includegraphics[angle=270,width=0.5\textwidth]{Fig2.ps}\\
\hspace*{-6mm}
\includegraphics[angle=270,width=0.5\textwidth]{Fig3.ps}
\caption{The total (top) and detectable (bottom) performance curves measured for every pre-processing method and choice of input parameters. The solid line is the reference performance curve measured from the unprocessed WSRT test datacube. The red X, blue crosses, green triangles and hollow circles mark the performance curves resulting from mathematical morphology subtraction, wavelet de-noising, linear smoothing and iterative median smoothing, respectively.}
\label{all_pcurves}
\end{figure}

The completeness-reliability performance curves that we measured for the various pre-processing methods are plotted in Figure \ref{all_pcurves}. We measured performance curves for all of the sources and for a subset of `detectable' objects, which are defined as having peak signal-to-noise ratios greater than or equal to three. We refer to these two types of performance curves as the total and detectable performance curves. The input parameters we used with these pre-processing methods are listed in Table \ref{params} in the Appendix. To obtain a meaningful comparison of the pre-processing methods, we used similar parameter values across the pre-processing methods. The choice of mathematical morphology opening subtraction kernels ranges from the the beam size to 20 times the beam size. This ensures that our choice of opening kernels brackets the optimal size of three times the source size, that was found by \citet{2002PASP..114..427R}.

Figure \ref{all_pcurves} shows that all of the pre-processing methods, except the mathematical morphology opening subtraction, produces a performance curve better than the default performance curve for the right choice of pre-processing parameters. The best performance curves are obtained by using iterative median smoothing or linear smoothing. Every method except for the linear smoothing can prove detrimental however if incorrect parameters are chosen. The iterative median smoothing has the most detrimental effect when a 5,5 kernel is used. This is as expected, because the kernel is larger than some source components. This is consistent with our observation that iterative median smoothing is always an improvement when the kernel's spatial extent matches the beam size. Conversely, as the mathematical morphology subtraction is more detrimental as the kernel size shrinks. We conjecture that the mathematical morphology subtraction performance curve is asymptoting towards the reference performance curve as the kernel approaches the size of the test datacube.

\begin{figure}
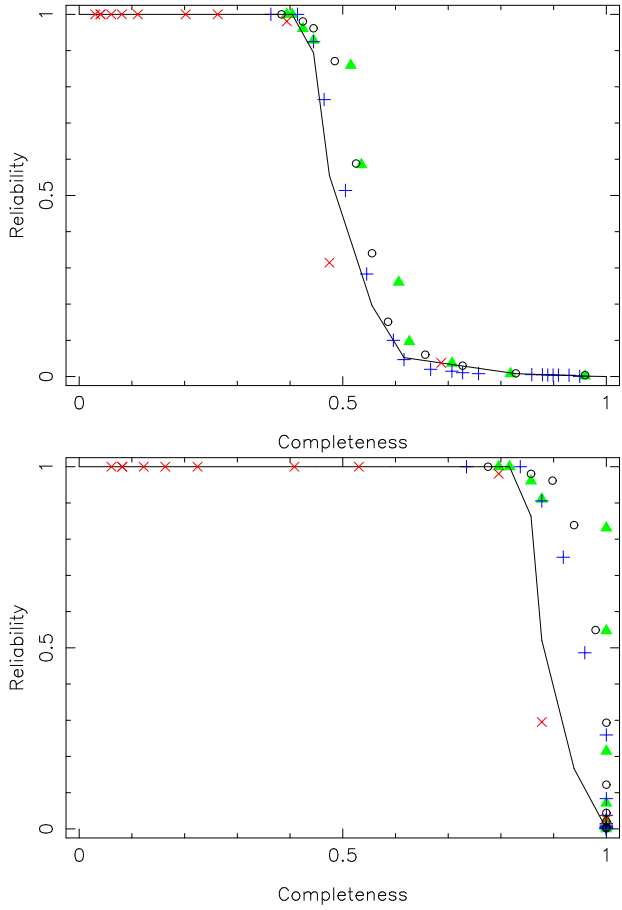

\hspace*{-6mm}
\includegraphics[angle=270,width=0.5\textwidth]{Fig4.ps}\\
\hspace*{-6mm}
\includegraphics[angle=270,width=0.5\textwidth]{Fig5.ps}
\caption{A subset of the total (top) and detectable (bottom) performance curves in Figure \ref{all_pcurves} (symbols have the same meaning). These performance curves are the best performance curves obtained with each pre-processing method.}
\label{pcurves}
\end{figure}

In Figure \ref{pcurves}, we have plotted a subset of the performance curves in Figure \ref{all_pcurves}, which reflect the best performance curve generated by each pre-processing method. The best results are achieved for: linear smoothing with an 11 channel Hanning filter in frequency; mathematical morphology opening subtraction with a single channel rectangle whose side length is 61 voxels; iterative median smoothing with a 3 by 3 by 11 rectangular parallelepiped and wavelet reconstruction in either 1 or 3 dimensions with a 4-$\sigma$ threshold, without applying a minimum scale threshold. In the rest of our analysis, we will only use the wavelet reconstruction in one dimension to avoid redundancy. The parameters for these performance curves are highlighted in bold italics in Table \ref{params}. 

From the performance curves in Figure \ref{pcurves} we can draw four interesting conclusions. First, a datacube needs to contain sufficient large scale structure for mathematical morphology opening subtraction to be worthwhile, and our real world test datacube does not. Second, for our choice of parameters, the linear smoothing and iterative median smoothing produce better performance curves than {\sc Duchamp}'s 3-D wavelet de-noising. Third, the detectable performance curves show greater improvement than the total performance curves, because pre-processing is more beneficial for sources that are easier to detect. Our final conclusion is that smoothing along the frequency axis contributes more to improvement in the performance curves than smoothing spatially. Visual inspection of our test datacube reveals that most of our sources are marginally resolved spatially, but well resolved spectrally. This means that a filter `matching' our sources distribution is providing the biggest improvement to the performance curves, which is what we would expect in the presence of Gaussian noise. We can therefore conclude that our real world test datacube's noise is sufficiently Gaussian that matched filtering is the optimal method for source extraction. 

\begin{figure}
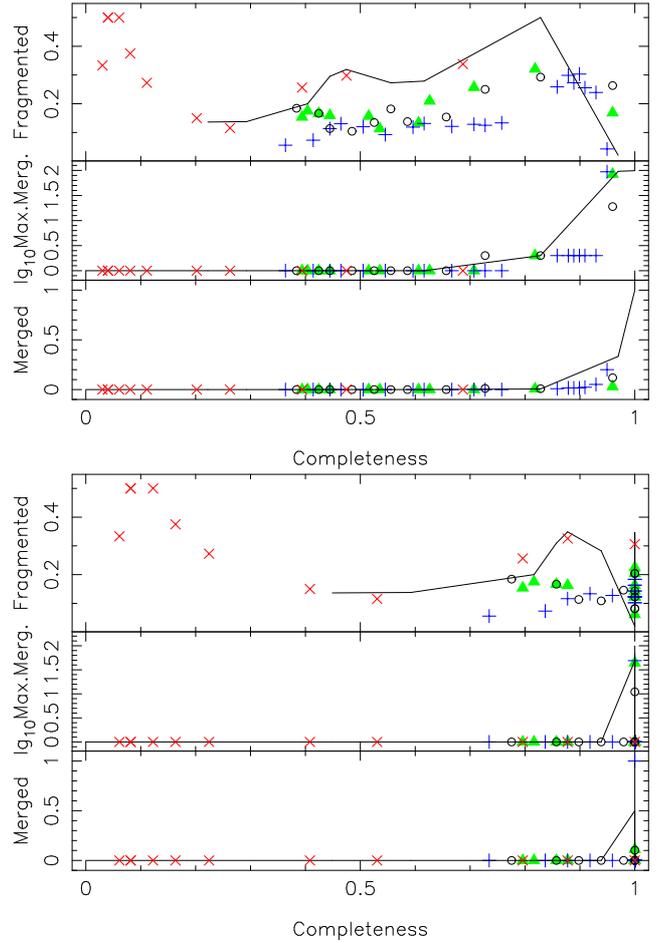

\hspace*{-6mm}
\includegraphics[angle=270,width=0.52\textwidth]{Fig6.ps}\\
\hspace*{-6mm}
\includegraphics[angle=270,width=0.52\textwidth]{Fig7.ps}
\caption{The merging rates, fragmentation rates and number of sources comprising maximally merged object for the total (top) and detectable (bottom) performance curve subset of Figure \ref{pcurves}, plotted against completeness. Symbols have the same meaning as Figures \ref{all_pcurves} and \ref{pcurves}.}
\label{ext_pcurves}
\end{figure}

We will further analyse the subset of pre-processing results in Figure \ref{pcurves}, by comparing the effects of the different pre-processing methods on the fragmentation (fraction of multiply detected sources) and merging (fraction of sources detected as a single object) rates as well as the number of sources contributing to the maximally merged object. The fragmentation and merging rates and maximally merged object are plotted in Figure \ref{ext_pcurves} as a function of completeness. 

Figure \ref{ext_pcurves} reveals that at the highest completeness values, the completeness comes at the expense of not only reduced reliability, but also an increasing merging rate. Eventually the merging becomes so bad that most of the sources are merged into a single object. This problem affects the linear smoothing the most, but it is no worse than the merging affecting the unprocessed datacube. The iterative median pre-processing produces the least merged source finding results at the highest completeness values. 

The wavelet de-noising produces the best (lowest) fragmentation rates at all completeness levels, but only marginally better than the linear smoothing and iterative median smoothing. Note that we have excluded the fragmentation rates at the highest completeness levels in our assessment, because the fragmentation rates decline due to increasing merging rates. The increase in fragmentation rates at low completeness levels (seen in the mathematical morphology subtraction) is expected when using an intensity threshold source finder. When using increasingly higher thresholds, only the peaks of a source are `found', which results in fragmentation. This fragmentation effect at low completeness levels is solved through the use of a `growth' threshold in addition to the threshold used to define source voxels.

\section{Conclusion}
\label{conclusion}
We have used the WSRT test datacube of \citet{2011arXiv1112.3162S} to test the real world performance improvement of linear smoothing, {\sc Duchamp}'s 3-D wavelet de-noising, iterative median smoothing and mathematical morphology subtraction, when using intensity thresholding to find sources in ASKAP HI spectral line observations. We generated completeness-reliability performance curves for each pre-processing method and an unprocessed datacube, which we used as a reference, to investigate the effect of each pre-processing method for a range of input parameters.

We found that the iterative median smoothing and linear smoothing produce the greatest improvement in source finder performance. We recommend that iterative median smoothing be used over linear smoothing though, because iterative median smoothing is less affected by merging and fragmentation. In our tests however the effect of iterative median smoothing on source finder performance proved to be more highly dependent upon the pre-processing parameters than the other methods. The performance improvement offered by {\sc Duchamp}'s 3-D wavelet de-noising was the least sensitive to the choice of input parameters. It is the safest pre-processing method. It should be noted however that using a smoothing kernel with a spatial extent smaller than or equal to the datacube's beam size, will in general improve source finder performance. 

We think the Gaussian nature of the WSRT noise is the reason that the linear smoothing and iterative median smoothing produce the greatest source finder performance improvement, because it approximates matched filtering. For this reason, we do not expect our results to be applicable to images or datacubes with non-Gaussian noise. We do however think that the edge-preserving nature of the iterative median transform is the reason that it does not suffer from fragmentation and merging as badly as linear smoothing, and that this result is applicable to all images and datacubes.




\appendix

\section{Ancillary tables}

\begin{table}
\caption{Performance curves were measured for the pre-processing methods using the combinations of input parameters in this table. Every combination of parameters in a given row has been used. All kernel sizes are given in voxels. The parameter combinations that produced the best performance curve for each pre-processing method (plotted in Figure \ref{pcurves}), are highlighted in bold italics.}
\label{params}
\vspace{1mm}
\begin{center}

\vspace*{-3mm}

Linear smoothing
\begin{tabular}{ccc}
\hline
smoothing axis & kernel & kernel size \\
\hline
\textbf{\emph{frequency}} & \textbf{\emph{hanning}} & 3, 7, \textbf{\emph{11}} \\
RA \& Dec & Gaussian & 3, 5 \\
\hline
\end{tabular}
\vspace{3.5mm}

Wavelet de-noising
\begin{tabular}{ccc}
\hline
dimensionality & min. scale kept & threshold \\
	& (voxels) & (sigma) \\
\hline
\textbf{\emph{1}}, \textbf{\emph{3}}  & \textbf{\emph{1}}, 3 & 2, \textbf{\emph{4}} \\
\hline
\end{tabular}
\vspace{3.5mm}

Iterative median smoothing
\begin{tabular}{ccc}
\hline
kernel & angular size & freq. size \\
\hline
\textbf{\emph{rect. parallel.}} & \textbf{\emph{3}}, 5, 7 & 3, 5, 7, \textbf{\emph{11}} \\
 & 11 & 11 \\
ellipsoid & 3, 5, 7 & 3, 5, 7, 11 \\
 & 11 & 11 \\
\hline
\end{tabular}
\vspace{3.5mm}

Mathematical morphology subtraction
\begin{tabular}{ccc}
\hline
kernel & angular size & freq. size \\
\hline
\textbf{\emph{rect. parallel.}} & 3 & 1, 3 \\
 & 5 & 1, 5 \\
 & 11 & 1, 11 \\
 & 21 & 1, 21 \\
 & 41, \textbf{\emph{61}} & \textbf{\emph{1}} \\
\hline
\end{tabular}

\end{center}
\end{table}

\end{document}